\documentclass[twocolumn,showpacs,preprintnumbers,amsmath,amssymb]{revtex4}

\usepackage{graphicx}
\usepackage{dcolumn}
\usepackage{bm}
\def\be{\begin{equation}}
\def\ee{\end{equation}}
\def\bee{\begin{eqnarray}}
\def\eee{\end{eqnarray}}

\begin{document}

\preprint{ }

\title{SGR 1806-20 and the gravitational wave detectors EXPLORER and NAUTILUS}

\author{G. Modestino}

\author{G. Pizzella}%
\email{giuseppina.modestino@lnf.infn.it; guido.pizzella@lnf.infn.it}
\affiliation{%
INFN, Laboratori Nazionali di Frascati, I-00044, Frascati (Roma) Italy\\
}
\title{SGR 1806-20 and the gravitational wave detectors EXPLORER and NAUTILUS}
\begin{abstract}
The activity of the soft gamma ray repeater SGR 1806-20 is studied in correlation with the EXPLORER and NAUTILUS data, during the year 2004, for gravitational wave (GW) short signal search. 
Corresponding to the most significant triggers, the bright outburst on October $5^{th}$ and the giant flare (GF) on December $27^{th}$, the associated GW signature is searched. Two methods are employed for processing the data. With the average-modulus algorithm, the presence of short pulses with energy $E_{gw}\ge1.8\cdot 10^{49}~erg$ is excluded with 90\% probability, under the hypothesis  of  isotropic emission. This value is comparable to the upper limits obtained by LIGO regarding similar sources. Using the cross-correlation method, we find a discrepancy from the null-hypothesis of the order of $1\%$. This statistical excess is not sufficient to claim a systematic association between the gravitational and the electromagnetic radiations, because the estimated GW upper limits are yet several orders of magnitude far away from the theoretically predicted levels, at least three for the most powerful SGR flare.
\end{abstract}

\maketitle

\section{Introduction}
Discovered since 1979 \cite{maze,helfa}, despite their unpredictable behavior, the soft gamma repeaters (SGRs) are considered interesting targets in gravitational wave (GW) studies. The fundamental ideas are the neutron star (NS) nature of these objects and the association of SGRs with magnetars \cite{duncan1,duncan2}. Detected as persistent X-ray source at $\sim 10^{35} $ergs $ s^{-1}$, they occasionally emit energetic soft gamma bursts,  up to $\sim 10^{42} $ergs $ s^{-1}$, or even much more energetic events. The large number of observed characteristics of SGRs, including the bursting activity during the three giant flares (GFs) detected to date \cite{maze,cline,mere,hurley2}, confirm the NS nature of  SGR, and offer an effective evidence of the presence of very high magnetic field ($B\sim 10^{15}$G). (For a review, see \cite{mere2}). According to several models, the burst triggers are primary reference studying the gravitational radiation \cite{freite,ioka,horv,owen,horo}.
 The most optimistic GW emission model \cite{ioka} foresees an extreme bursting activity, probably due to a sudden global reconfiguration of the internal magnetic field, and the magnetar shape deformation, causing the increase in the momentum of inertia with a release of energy through gravitational radiation. 
Also the proximity ($\sim10$ kpc) makes these objects intriguing for GW searches, enhancing chances of detectability. 

To this aim, several measurements had previously been performed. Studying  the GF occurring on December  $ 27^{th} $ 2004, the AURIGA group \cite{cerdonio} explored the frequency range 930-935 Hz, under the hypothesis of oscillating emission with a damping time of 100 ms. Expressing the result in terms of the dimensionless amplitude $h$, they found an upper limit of  $ 2.7\cdot 10^{-20}$, at the time of the hyperflare. In relation to the same event, the LIGO collaboration  \cite{abbott} examined the pulsating tail of the burst which revealed the presence of quasi-periodic oscillations (QPOs) in the X-ray light curve, as RXTE and RHESSI satellite had detected \cite{watts,israel}. LIGO found no excess and, racing from $\sim 10^{47}$ ergs, they set several upper limit levels on the GW emitted energy, depending on time and frequency radiation. More recently, the same LIGO collaboration presented the results of short-duration GW events associated with SGR 1806-20 and SGR 1900+14 storms  occurred during the year starting from November 2005 \cite{abbott1,ligo3,kalmo}. Including the GF on December $27^{th}$, they analyzed almost two hundred  events finding no evidence of any association. Depending on signal waveform types, several upper limit are estimated, referring to more than five orders  of the GW energy magnitude (from $10^{45}$ erg to almost $10^{51}$ erg) for hypothetical isotropic GW emission.
	
In the present paper we consider the particular phenomenon of the SGR 1806-20 during the year 2004.  The  resonant GW detectors EXPLORER and NAUTILUS \cite{asto3} data are analyzed corresponding to the long sequence that occurred on October $5^{th}$ \cite{gotz}, and to the GF on December $27^{th}$. As previously seen, during this event, several observations were performed using GW experimental apparatuses  but, as far as the October outburst is concerned, there were no previous GW measurements although it represents a very significant event because of the exceptional energy release and duration, extending $3\cdot 10^{42}$ erg over more than one hundred pulses in a ten minutes span of time. 

The principal points of the present analysis are as follow. The presence of  short  pulses (up to a few tens of milliseconds) was searched for on both GW detectors, using the data filtered by  linear  filter matched to $\delta$-like signals. In Sec.\ref{sensi}, the sensitivity of the apparatuses is shown relatively to the year 2004. As explained in Sec.\ref{onsource}, the physical regions were chosen on two long sequences of twenty-six plus forty-six seconds, corresponding to the highest EM peaks of the October $5^{th}$  outburst, and on one second centered around the trigger time of the GF on December $27^{th}$. The long period, in the case of the outburst, and knowledge of the arrival time on Earth of a light-speed signal at the time of the GF \cite{hurley2,cerdonio} are circumstances particularly opportune to avoid uncertainty in the analysis focusing on the reference time of the gravitational radiation. The expected distribution under the null-hypothesis is built using real data, with random time shifts  of one GW detector with respect the other one (Sec.\ref{offsource}). Essentially, two algorithms are employed. With the first, discussed in Sec.\ref{algo1}, we evaluate the excess amplitude on the on-source regions of the GW detector data. With this measurement (calibrated using the cosmic ray signals \cite{cosmici} , as explained in the Appendix), we exclude the presence of GW short signal and we establish the upper limit to the amount of isotropic GW energy $E_{GW}$ emitted during the exceptional periods. The second algorithm regards the cross-correlation function calculated between the EXPLORER and NAUTILUS data streams, as shown in Sec.\ref{algo2}. Following previous methods \cite{asto1,asto2,ligo1,ligo2}, cumulative analyses are performed on the sample of seventy-three measurements obtaining several mismatches of the order of $1\%$ probability or less, from the background. The statistical significance is evaluated on the basis of the local probability and binomial tests of the resulting loudest measurements.
In the final discussion, Sec.\ref{discu}, we provide a summary of the study and further comments.

\section{EXPLORER and NAUTILUS sensitivity}
\label{sensi}
We consider the EXPLORER and NAUTILUS operating during 2004 \cite{asto3}, using the filtered data set-up as the ROG collaboration released \cite{esperirog}. The main characteristics of the experimental apparatuses, in terms of frequency coverage, are shown in table \ref{cara}. The frequency region where the detectors are sensitive covers a range of about one hundred hertz, thanks to which the data can be sampled at time intervals of 3.2 ms. The data are filtered with a filter matched to delta-like signals for the detection of short bursts \cite{fast},  typically of the order of a few milliseconds.

Let $x(t)$ be the filtered output of the detector. 
For well behaved noise due to the thermal motion of the oscillators and to the electronic noise of the amplifier, the distribution of $x(t)$ is normal with zero mean. Its variance represents the $effective~temperature$ and is indicated with $T_{eff}$. The distribution of $x(t)$ is
\be
f(x)=\frac{1}{\sqrt{2\pi T_{eff}}}e^{-\frac{x^2}{2T_{eff}}}
\label{normal}
\ee
 During the year 2004 $T_{eff}$ was 2.0 mK for EXPLORER and 1.7 mK for NAUTILUS.
 
\begin{table}
\centering
\caption{Resonant frequencies of EXPLORER and NAUTILUS during 2004, instrumental bandwidth $\Delta f$ and relative sensitivities.}
\vskip 0.1 in
\begin{tabular}{|cccc|}
\hline
detector&resonances $f [Hz] $& $\Delta f [Hz]$&$H_{min}[\frac{1}{Hz}$]\\
&&&\\
\hline
 EXPLORER&904.7,   921.3 &8.7&$3.4\cdot10^{-22}$\\
 NAUTILUS&926.3,   941.5 &9.6&$3.1\cdot10^{-22}$\\
\hline
\end{tabular}
\label{cara}
\end{table}

The resonant detectors are able to measure the Fourier component $H$ of the incoming gravitational radiation with adimensional amplitude $h(t)$. The apparatuses sensitivity (SNR=1) to short burst can be expressed as the minimum detectable $H$ \cite{amaldi,naut}
\be
H_{min}=\frac{1}{4Lf_0^2}  \sqrt{ \frac{k_{B}T_{eff}}{M}}~~~ [\frac{1}{Hz}]
\label{Hmin}
\ee
where $k_{B}$ is the Boltzmann constant and $f_0$, $L$ and $M$ are the resonance frequency, the length and the mass of the bar.
In general, if we detect a burst with energy  $E_{s}$, expressed in kelvin units, the corresponding value is
\be
H=\frac{1}{4Lf_0^2}  \sqrt{ \frac{k_{B}E_{s}}{M}}\simeq 2.4 \cdot10^{-22}\sqrt{E_s(mK)}~~~ [\frac{1}{Hz}]
\label{HH}
\ee
for optimal orientation.

Besides the single bursts, this search is suitable for any GW transient which shows a nearly flat Fourier spectrum at the two resonant frequencies of each detector. The metric perturbation $h(t)$ can either be a millisecond pulse, a signal made by a few millisecond cycles, or a damped sinusoid signal. In the hypothesis of a signals sweeping in frequency through the detector resonances, with small decay times ($< 50~ms$), the filter maintains good detection capability \cite{pai1,pai,igec}.
For a short signal of duration time $\tau_g$ and frequency $f_g\sim f_0$,
the spectral amplitude, $\tilde{h}=\sqrt{\int{|H(f)|^2~df}}$, can be put
\be
\tilde{h}
\simeq
H\sqrt{\pi /\tau_g}~~~ [\sqrt{\frac{1}{Hz}}]
\label{spectra}
\ee
 
\section{Data selection}
\label{data}
During the year 2004 several detectors on spacecrafts observed a special activity from SGR 1806-20 \cite{nasa} culminating on December $27^{th}$. Before the GF, other considerable events occurred, such as the strong outburst October $5^{th}$, when more than one hundred short bursts were emitted in a few minutes, involving more than $3~10^{42}$ erg in terms of EM energy release \cite{gotz}. In the last reference, there is an accurate description of the event and relative light curves of the initial brighter part. The instruments (in the 15-200 keV energy range) indicate two very intense peak clusters, the first one in the vicinity of the trigger at 13:56:49 UT, the second one after about three minutes.

\subsection{On-source region}
\label{onsource}
Referring to the cited  EM triggers, we consider physically interesting the following time periods of the filtered data of both EXPLORER and NAUTILUS: 

- $\it{a)}$ on October $5^{th}$, 13:56:44 - 13:57:10, 26s (starting 5s before the EM trigger); 

- $\it{b)}$ on October $5^{th}$, 13:59:39 - 14:00:25, 46s during the second cluster of bright bursts;

- $\it{c)}$ on December $27^{th}$, 21:30:26.14 - 21:30:27.14, 1s centered to the GF trigger time \cite{hurley2,mere}.

The choice of the two long first periods is due to the special profile of the light curves, but it is also opportune to avoid any uncertainty about the time delay between the two emissions that make the time reference definition difficult in the correlation study. Nevertheless, relatively to the GF, it is possible to restrict the measurement period because the impinging time of a light speed signal onto the GW detectors is known \cite{cerdonio}.
Since the three different lengths (26s, 46s, 1s) don't allow to employ an easy algorithm for a uniform analysis from the statistical and physical point of view, we divide the two long periods into a number of contiguous intervals, 1-second each one, obtaining a comprehensive sample of seventy-three independent segments for performing measurements. We call the GW detector data in this region on-source data. All routine checks are done on the selected period, principally by vetoing with a threshold on $T_{eff}$, to insure good sensitivity and stationary data. In addition, we also control the presence of local noise anomalies, and the effectiveness of the division into 1-second segments, evaluating the autocorrelation function to the EXPLORER and NAUTILUS data of the on-source  $\it{a}$ and $\it{b}$ periods, as explained in the Appendix. 
\subsection{The background data}
\label{offsource}
For estimating the expected result under the null-hypothesis, we use real data but considering time regions distant a few hours away the EM triggers. By assigning 3000 random time steps, we obtain as many off-source 1s segments, for the first GW detector. To avoid simultaneous signals of any nature, we apply a time shift of several hours on the data sequence of second GW detector with the respect the first one, thereby obtaining as many as 3000 off-source segments. All periods are selected vetoing by a threshold on $T_{eff}$, consistently with the on-source region selection.
\section{The average algorithm}
\label{algo1}
The first algorithm for scanning the on-source data is designed to detect an enhancement due to a GW burst.
Due to the oscillating character of the response, we average over $1s$ the amplitude modulus of each sample, $x_{i}$ for EXPLORER and $y_{i}$ for NAUTILUS, belonging to the on source segments.  The averaged signal parameter is indicated with 
\be
AS \equiv  \frac{\Sigma_{i=1}^{n} |x_{i}|+\Sigma_{i=1}^{n} |y_{i}| }{2n}~~~~[\sqrt{K}]
\label{as_par}
\ee
where $n=313$, the number of 3.2 ms samples in each second.
 \begin{figure}
\includegraphics[width=1.0\linewidth]{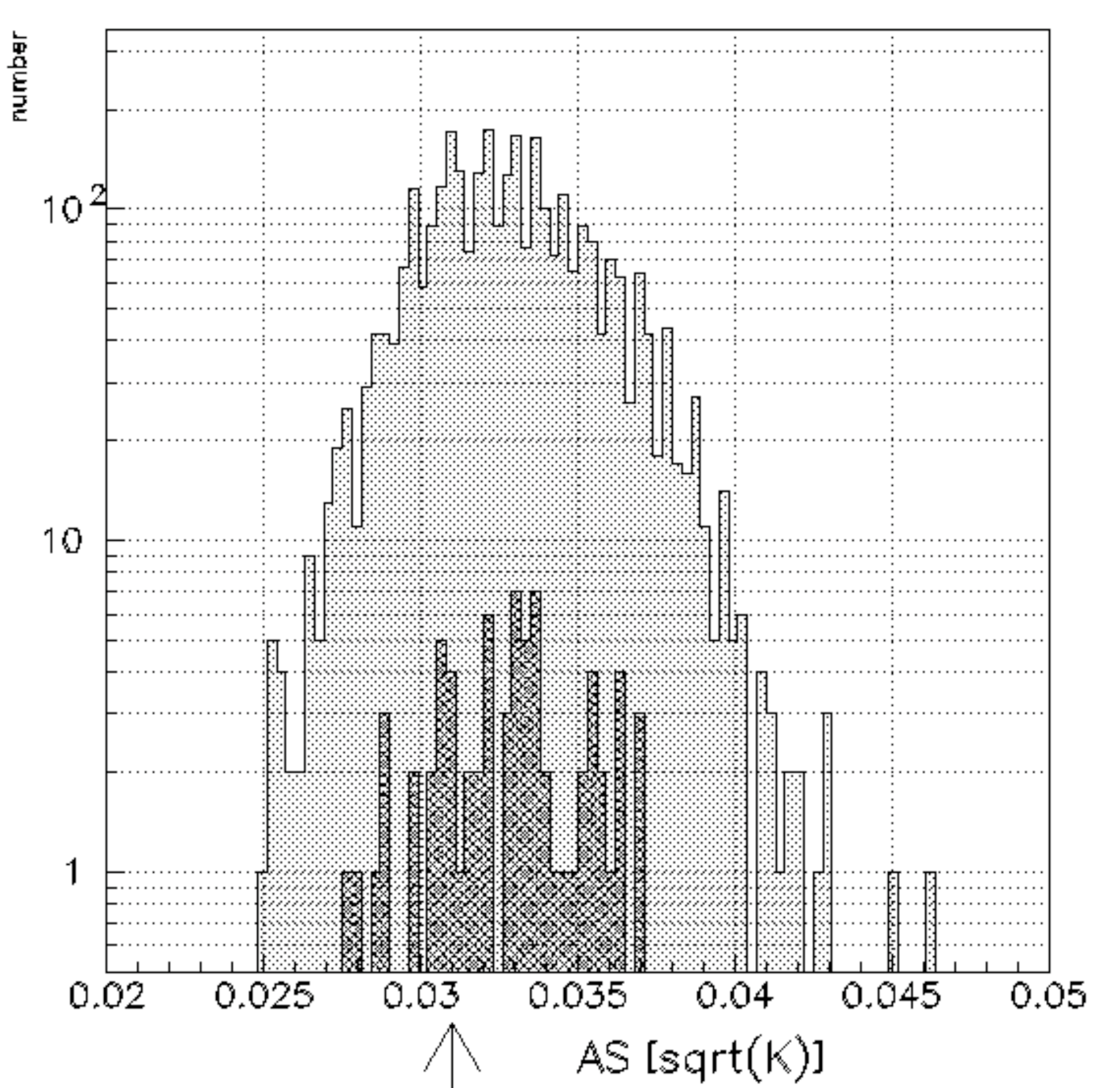}
\caption{
Distribution of the 73 AS parameter values (the darkest area), as defined in Sec.\ref{algo1}. The lightest part is the background distribution of the same parameter evaluated on the off-source regions. The Kolmogoroff compatibility test gives 45\% probability. The arrows indicates the AS evaluation on the data streams at the time of the GF.
\label{fig1_as_distri} }
\end{figure}
Applying the eq.\eqref{as_par} to the 73 pairs on-source segments, we obtain the distribution in fig.\ref{fig1_as_distri}. The corresponding background, the lightest area of fig.\ref{fig1_as_distri}, is calculated on the 3000x2 segments that have previously defined. Cumulative calculations give:
$\bar{AS}_{73}=3.29\cdot10^{-2}~\sqrt{K}$, very close to the $AS_{bck}$ value ($3.28\cdot10^{-2}$, with root-mean-squared, $RMS=2.8\cdot10^{-3}~\sqrt{K}$).
Comparing the distributions by the Kolmogoroff test, we obtain probability 45\% that the AS measured distribution be similar to the background one.
The high compatibility value and, as we can easily note, the absence of loud events in the distribution, clearly indicate that no amplitude excess is present on the physical region.
Thus, we don't retain to perform further statistical investigation on this result. Rather, considering the particular
phenomenon and the experimental calibration by the CR excitation, this measure assumes an important significance in terms of upper limit for energy evaluation of emitted gravitational radiation.
\subsection{Upper limit}
\label{ulimit}
In the present study, two different levels are present in terms of source type. 1) Several peaks belonging to the outburst on October $5^{th}$,
 when the EM energy release of $3\cdot10^{42}$ erg was involved. 2) The GF with of $1.6\cdot10^{46}$ erg of isotropic $E_{EM}$ in the 
 main peak \cite{hurley2}. There are four order of magnitude in terms of emitted energy between them, so we need to distinguish two measurement levels. In Tab. II, corresponding to each category, we set out the AS measured parameter. By the cosmic ray calibration procedure, as explained in appendix, the input signal energy values are extracted. From the eqs.\ref{HH} and \ref{spectra}, assuming an incoming burst with duration time $\tau_g $ and $f\sim f_0$, the spectral amplitudes $\tilde{h}$  are evaluated. Under the frequentist approach \cite{feco}, the corresponding upper limits are calculated, at 90\% confidence level, setting  $\tilde{h}=2.9\cdot10^{-21}~\sqrt{1/Hz}$, for the period $1)$, and 
 $\tilde{h}=4.3 \cdot10^{-21}~\sqrt{1/Hz}$, for the period $2)$.
 
 Using the classical resonant detector cross-section, under the hypothesis of isotropic emission, the total energy carried by the gravitational waves is given by
\be
E_{GW}=\frac{\pi^2 c^3}{8G}~\frac{1}{ML^2f^2}\frac{r^2}{\tau_g}E_s=
\nonumber
\ee

\be
=37.4\cdot10^{49}\frac{1~ms}{\tau_{g}}(\frac{r}{10~kpc})^2E_s(mK)~~~[erg]
\label{egw}
\ee
where $r$ is the source distance, $\tau_{g}$ is the signal duration and $E_s$ expressed in mK units. For the categories in Tab.\ref{primatavola}, under the hypotheses of an isotropic emission, the corresponding $E_{GW}$ upper limits are $1.8\cdot10^{49} ~erg$ and $4\cdot10^{49}~erg$,
 with duration time $\tau_{g}=30ms$ and $r=10~kpc $. In order to compare
  the results to previous similar measurements, we need to consider the ratio $\gamma=\frac{E_{GW}}{E_{EM}}$.
The importance of this parameter is well pointed in
a series of LIGO analyses regarding the association with SGRs \cite{abbott1,ligo3,kalmo}.
Their results are distributed over a wide
interval comprising  almost eight orders of magnitude in terms of $\gamma$, depending on several signal waveform hypotheses and on relative
sensitivities.
In our measurements, $\gamma^{90\%} =6\cdot10^{6}$, for October outburst, and $\gamma^{90\%} =2.5\cdot10^{3}$, for the December GF. These values  are included in the ranges established by previous cited measurements.
\begin{figure}. 
\includegraphics[width=1.0\linewidth]{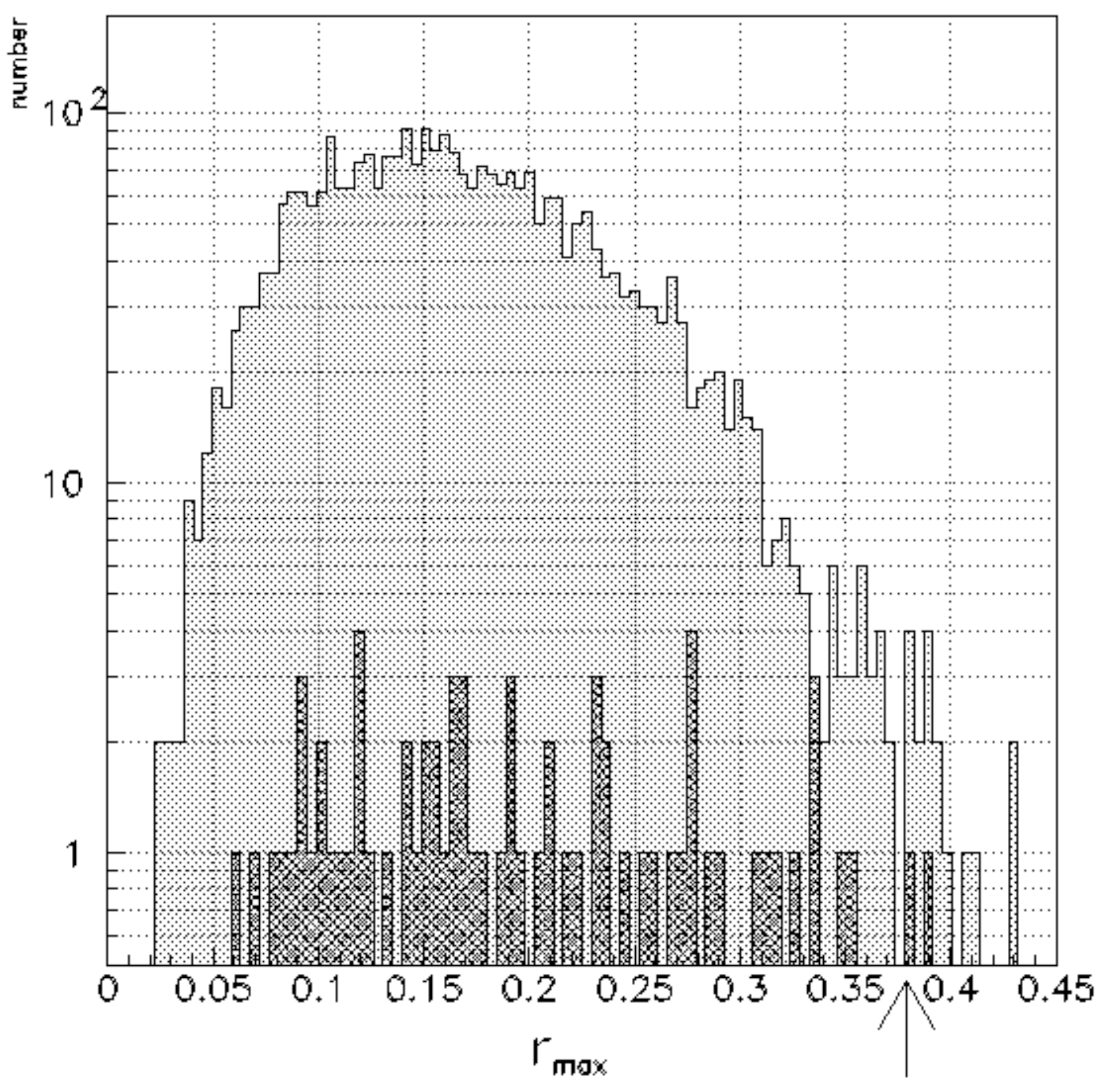}
\caption{
Distribution of $r_{max}$, the maximum value of the cross-correlation function for $\tau=\pm16$ ms, evaluated on the seventy-three on-source regions (the darkest area). The lightest part is the distribution of the same parameter evaluated on the background 3000 off-source regions. The Kolmogoroff compatibility test gives 1.3\% probability. The arrows indicates the  value at the time of GF. 
\label{fig2_rmax_distri} }
\end{figure}

\begin{table}
\centering
\caption{Upper limits for gravitational waves with the EXPLORER and NAUTILUs detectors, using the AS quantity. }
\vskip 0.1 in
\begin{tabular}{|c|c|c|c|c|}
\hline
period&$AS$&$\tilde{h}^{ 90\%}$&$E_{GW}^{90\%}$&EM\\
2004 day~~~~lenght(s)&$[\sqrt{K}]$&$[\frac{1}{\sqrt{Hz}}]$&[erg]&[erg]\\
\hline
&&\\
1) Oct $5^{th}$~~~~~~26+46&$3.29\cdot 10^{-2}$&$2.9\cdot 10^{-21}$&$1.8\cdot10^{49}$&$~~3\cdot10^{42}$\\
& ${}^{\pm 0.3\cdot 10^{-3}}$&&&\\
2) Dec $27^{th}~~~~~~~~1~~~$&$3.11 \cdot10^{-2}$&$4.3\cdot 10^{-21}$&$4\cdot10^{49}$&$1.6\cdot10^{46}$\\
& ${}^{ \pm 0.28 \cdot 10^{-2}}$&&&\\
\hline
\end{tabular}
\label{primatavola}
\end{table}

\section{The cross-correlation response}
\label{algo2}
We correlate the data of EXPLORER and NAUTILUS processed with the filter
matched to delta-like signals by taking simultaneous time periods of data of both
detectors lasting one second each (on-source regions) calculating
 the cross-correlation function
\be
r(\tau)=\frac{\varepsilon[(x(t+\tau)-\bar{x})(y(t+\tau)-\bar{y})]}{\sqrt{\varepsilon[(x(t)-\bar{x})^2\varepsilon[(y(t)-\bar{y})^2}]}
\label{cosr}
\ee
where the filtered data of EXPLORER are indicated with $x(t)$, and the filtered data of NAUTILUS, with $y(t)$. $\bar{x}$ is the average value of $x$, $\bar{y}$ that of $y$ and $\varepsilon[...]$ is the $expectation$. We consider the maximum value of the correlation function
\be
r_{max}\equiv max~r(\tau)~~~~\tau\in[-16ms,16ms]
\label{rmax}
\ee
In this way we take care of possible small time mismatches 
between the two detectors and also of the fact that the real excitation may not
be properly shaped with the delta-function used in the matched filter.
(In the Appendix, we show how this procedure is applied to data which include
cosmic ray excitations.) The distribution of the seventy-three $r_{max}$ values is shown in fig.\ref{fig2_rmax_distri} with the relative background.
Also for $r_{max}$ the expected distribution under the null-hypothesis  is empirically evaluated using the 3000 pair stretches built as described in Sec. III B.
The comparison by Kolmogoroff test gives $1.3\%$ of probability that the two distributions be similar.
 \begin{figure}
\includegraphics[width=1.\linewidth]{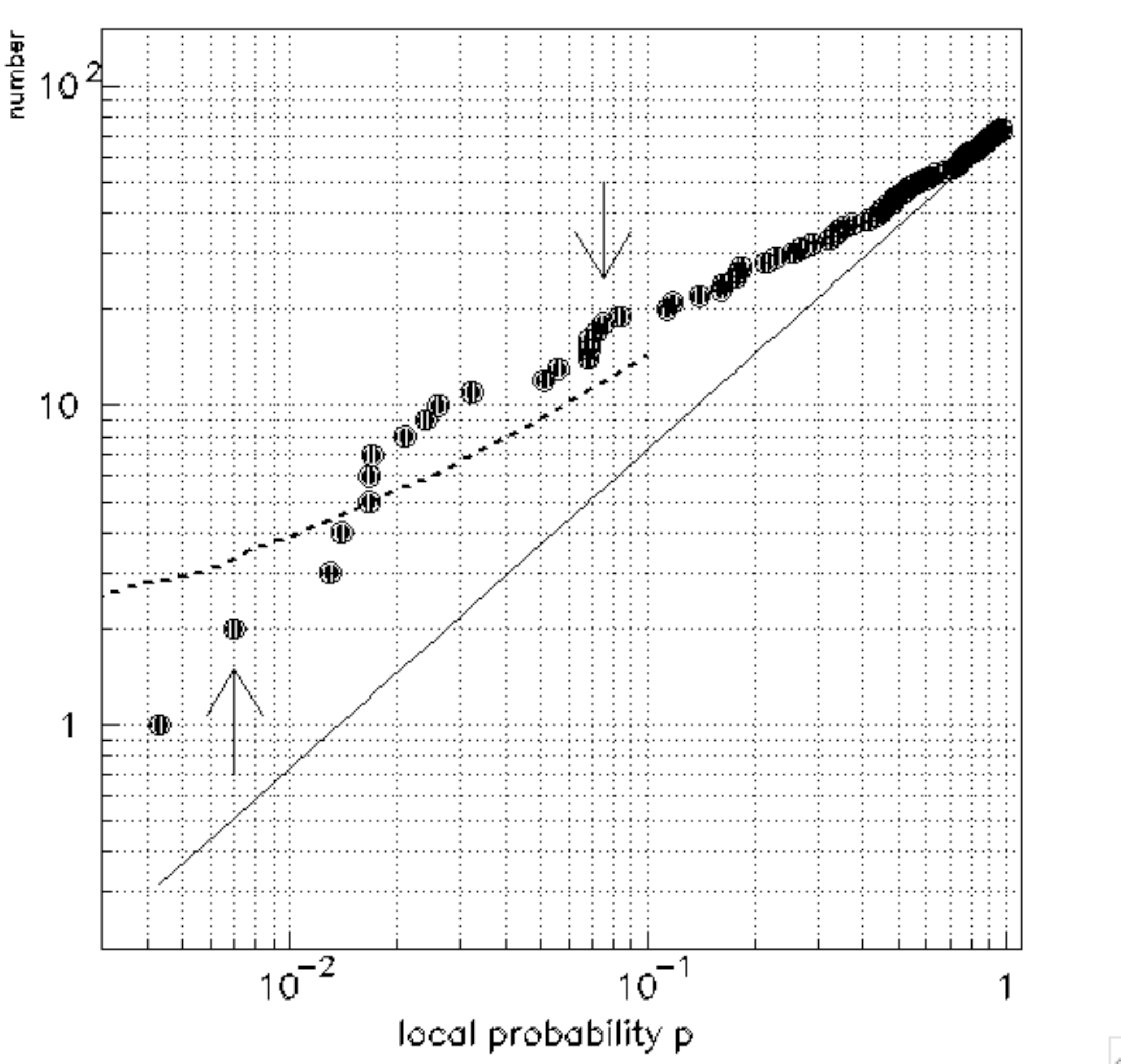}
\caption{
The cumulative local probability $p$ distribution for $r_{max}$ parameter. The seventy-three on-source measurements are represented by the bullets. The continuos line is the null-hypothesis $r_{max}$ distribution normalized to the number of measurements. The dashed line gives the excess needed for a $1\%$ confidence level in the null hypothesis. The up-arrow indicates the local probability level of $r_{max}$ at the time of the GF. 
The down-arrow points to the best real result in terms of binomial probability $P_{\geq 18}(0.075)=5.5\cdot10^{-6}$
 \label{fig3_rmax_pro} }
\end{figure}

\subsection{Statistical tests}
\label{sta_test}
Considering that GW signal from individual $1s$ time periods could be likely weak, we test for a cumulative signature associated with the whole segment sample.
In the past, this method has been applied to study correlations between EXPLORER and NAUTILUS data and gamma ray bursts (GRBs) \cite{asto1,asto2}.
It has been also used and developed in the second, third and fourth LIGO science run data analysis \cite{ligo1}, and more recently, in the enlarged LIGO-Virgo collaboration for GRB search \cite{ligo2}.
In the last cited papers, we find especially useful the probability distribution study and the binomial test, in order to better appraise the statistical significance of the loudest correlations and to select candidate events.
In our case, the same method can be helpful to mark time segments that particularly deviate from the background.
The cumulative local probability  distribution (fig.\ref{fig3_rmax_pro}) is associated to the $r_{max}$ measurements. The tail of the $p$ local distribution corresponds to the loudest $r_{max}$ values. In the same figure the expected distribution under the null hypothesis is indicated by the solid line. As easily noted, the  $p$ tail shows an extended deviation from the background. Corresponding to the measurements with local probability  $p\leq 0.1$, we calculate the binomial probabilities $P_{\geq i}(p_i)$ of getting $i$ or more events with $p\leq p_{i}$, obtaining  a spectrum which ranges down to $5.5 \cdot 10^{-6}$ (at $p_{18}=0.075$).
The measurement relative to the 
GF segment gives the second largest $r_{max}$ value, setting $p_2=7 \cdot 10^{-3}$, and  $P_{\geq2}= 0.093 $. We perform further statistical investigations taking into account unforeseeable non-stationary effects in the data, and the empirical form of the employed algorithm.
Sorting $k=20000$ different arrangements,  composing each one with 73 random draws from the null-hypothesis distribution of fig.\ref{fig2_rmax_distri}, we estimate the binomial value $P_{\geq i}
(p_{i},k)$, and the minimum ${P(k)}_{min}$, between the seventy-three values of the $k^{th}$ set. Comparing it with the real best result, we count 
how many times ${P(k)}_{min} \leq5.5\cdot10^{-6}$. It happens two times out of the 20000 different arrangements. We also calculate the expected number of events for having an 1\% confidence excess with respect the null hypothesis, at given probabilities $p\leq 0.1$. These values are shown in fig.\ref{fig3_rmax_pro} with the dashed line.
From the comparison with the $r_{max}$ on-source distribution (the bullets in the same figure), we note a discrepancy with the null-hypothesis with probabilities well below $1\%$, in the region $0.02\leq p\leq 0.1$.
\section{discussion}
\label{discu}
We studied the EXPLORER and NAUTILUS data in correlation with the astrophysical gamma-ray bursts during the year 2004, when an exceptional activity  was observed in terms of electromagnetic emission. In particular, we examined the long outburst that occurred on October, considering the two time intervals containing the major peaks, (respectively of twenty-six and forty-six seconds), and $1s$ time interval centered  on the outstanding flare on December $27^{th}$. On the resulting sample of seventy-three segments in total, $1s$ each one, we employed two algorithms. The first one consists in averaging the absolute values of the
EXPLORER and NAUTILUS responses on the on source region, and calibrating with the
transfer function of  real delta-signals due to the cosmic ray excitations. 
Assuming the hypothesis of GW short pulses, no significant excess is found either with the overall statistic sample, nor analyzing the segments with loudest amplitude values.
We obtain 
the upper limit expressed as amplitude signal or GW energy released by the astrophysical source with respect to two energy ranges.
Considering the bright outburst of October, involving a release of $\sim10^{42}$ erg, the result is $E_{GW}^{90\%}=1.8\cdot10^{49}$ erg. With regard to the GF, when an isotropic electromagnetic energy  of $1.6~10^{46}$ erg is assumed, the upper limit is $4\cdot10^{49}$ erg. They are comparable with previous values obtained by LIGO, in the frequency band 100-1000 Hz. 

We also processed the EXPLORER and NAUTILUS data, on the same on-source segments,  using the cross-correlation algorithm and applying a series of statistical tests. This analysis shows a compatibility of the order of $1\%$ with respect to the null hypothesis, for many on source segments. Considering the $\gamma$ factor, the parameter probing into the GW emission efficiency, we obtain $\sim10^{7}$ and $\sim10^{3}$, implying that the measure results are yet  far away from the expected level under theoretical hypotheses. Therefore, the use of statistical arguments alone is not sufficient for deducing a systematic association between the two phenomena, but the study can turn out useful for further investigations.
More observations are necessary in the direction of these intense sources. From this point of view, the recent improvements in gamma-astronomy are very encouraging. Over last few years, various other important detectors have been launched, such as the AGILE mission \cite{tavani} and Fermi Gamma-ray Space Telescope \cite{kaneko}. Other magnetars were also detected, such as the SGR 0501+4516 \cite{bart}, the SGR 0418+5729 \cite{van}, and the very recent SGR 1833-0832 \cite{bart1}. Thus, a fair amount of solid evidence exists for a more definite scientific strategy, studying multi-messenger phenomena.
\section{acknowledgements}
We gratefully acknowledge the ROG collaboration for having made available the data of the GW detectors, and for the useful discussions.

\section{Appendix}
\label{appe}
\subsection{Autocorrelation test}
\label{auto}
Performing the GW data analysis, three time intervals are chosen, as explained in the Sec.\ref{data}. The first two periods are a few ten seconds long each one, but they  have been divided  into 1-second segments for the analysis. To check the presence of local noise structures, as little glitch sequences or long bumps, that could invalidate the measurement also from the statistical point of view, we evaluated the autocorrelation function to the EXPLORER and NAUTILUS data of the on-source  $\it{a}$ and $\it{b}$ periods chosen in the Sec.\ref{data}.  The result is shown in the fig.\ref{fig0_auto}. Although the checks are performed using fifty seconds of data for each period, the graphics are presented up to one second of time shift $\tau$, given that the autocorrelation function drops within $\sim 200~ ms$ , as we easily note, and the same behaviors continue on the whole periods.

\subsection{Cosmic rays detection}
\label{cali1}
The test validity is performed employing the real signals generated by
extensive air showers (EASs) \cite{cosmici}, and the filtered data of EXPLORER and
NAUTILUS, all relative to the year 2004 \cite{asto3,esperirog}. Two events were
selected, each for the single detector, with highest interaction
energy with relative bar, in order to be independent of the noise. The responses of the two detectors are shown in fig.\ref{fig4_cos}. The
selected EASs have respectively 3561 and 3128 of particle multiplicity
measured in the lower part of the cosmic ray detectors. The vibrational energy
peaks are correspondently $2.47 K$ and $1.08 K$, several hundred times the
background, indicating a total energy of the order of $10~ TeV$ deposited in
each bar \cite{cosmici}.

\subsection{ Average algorithm calibration}
\label{fig0_auto}
We use the typical envelope (see fig.\ref{fig4_cos}) for simulating
delta-like signals. Obviously, we opportunely reduce each signal by scale factor
 depending on the choice of the amplitude signal. Then, we add the
constructed signal to the centre time of 300x2 off-source regions
using $10\%$ of the background employed in the Sec.\ref{offsource}. In terms of energy amplitude six values are chosen in the range $0.5-30 ~mK$, and for each level, the average algorithm is applied
obtaining the corresponding $AS$ responses applying the eq.\ref{as_par}. The calibration curve is reported in the fig.\ref{calibra}. At $E_s=0$, the $AS_{bck}$ value is reported, as calculated in Sec.\ref{algo1}.
 \begin{figure}
\includegraphics[width=1.0\linewidth]{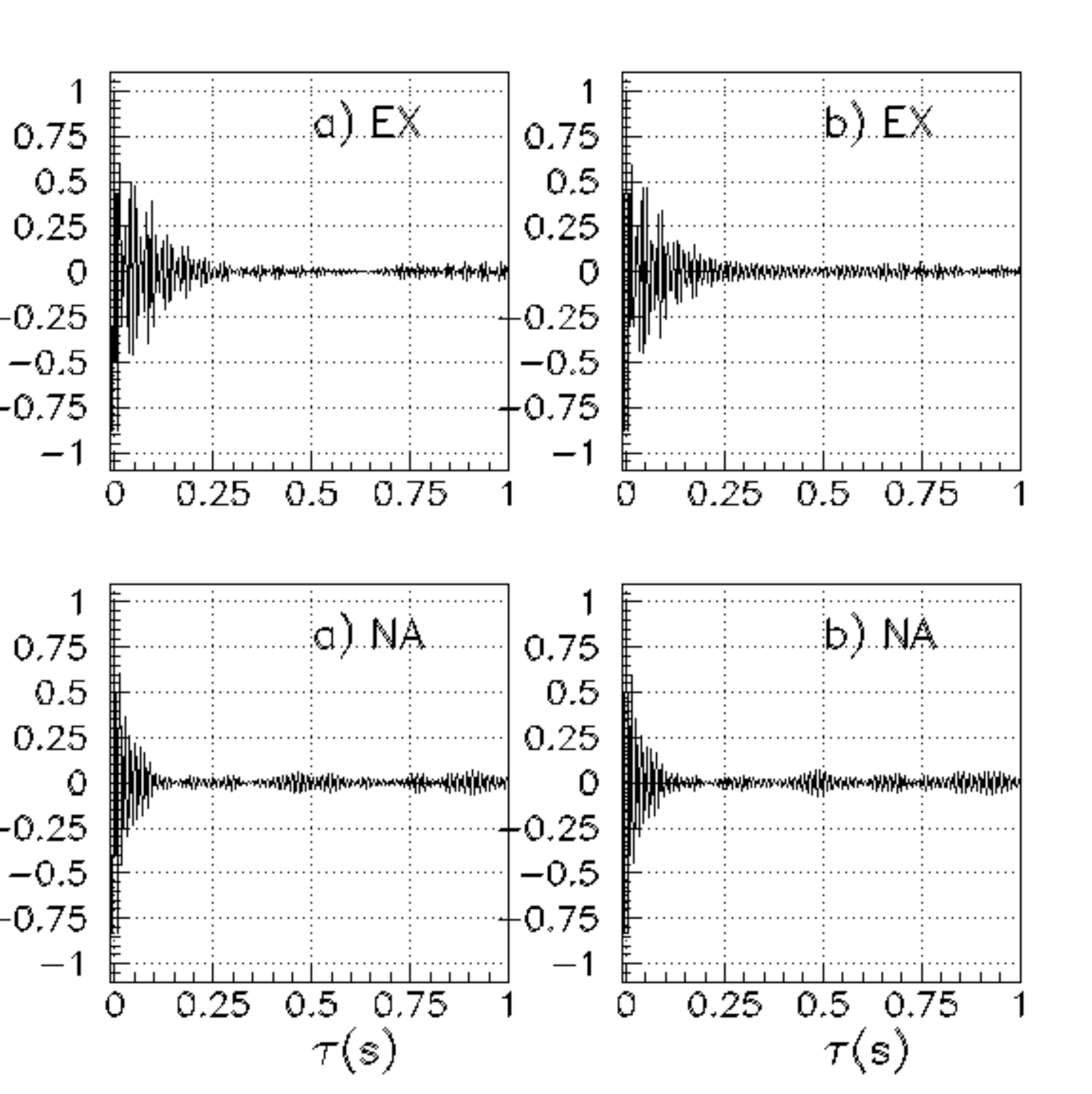}
\caption{
Autocorrelation functions for both EXPLORER (up) and NAUTILUS (down) data. The checks are performed using  fifty seconds  overlapping the on-source periods $\it{a}$ and $\it{b}$ (respectively on the left and on the right of the figure) both belonging to the October $5^{th}$ outburst.
\label{fig0_auto} }
\end{figure}

\begin{figure}
\includegraphics[width=0.9\linewidth]{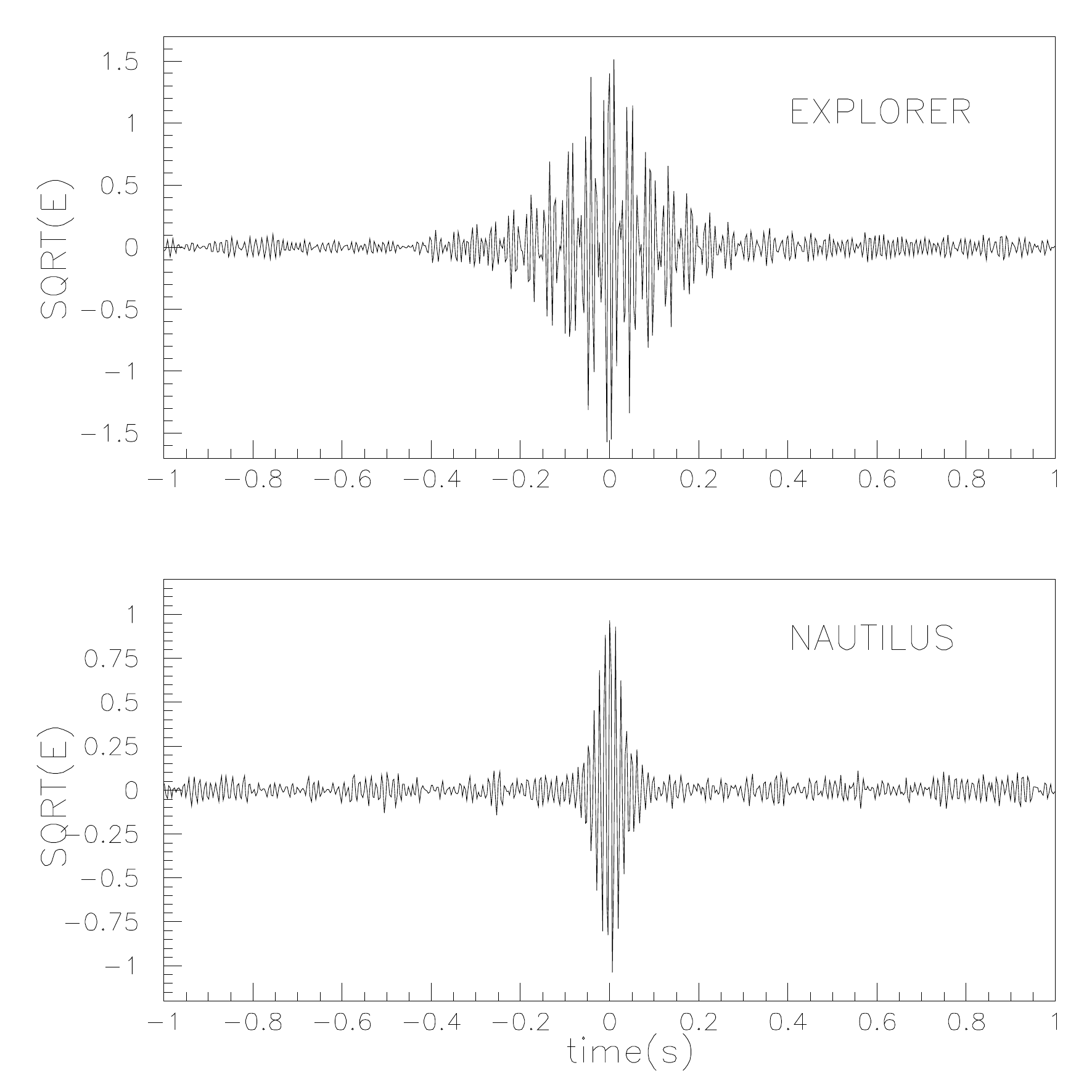}
\caption{
Two selected cosmic ray showers producing in the two detectors signals much larger than noise. On the ordinate axis we show the $\sqrt{E}$, with the energy $E$ in kelvin units. We notice the different resonance frequencies in the two detectors. The time of the two signals are shifted to an arbitrary $zero$, in order to simulate two simultaneous signals.
   \label{fig4_cos} }
\end{figure}

\begin{figure}
\includegraphics[width=0.9\linewidth]{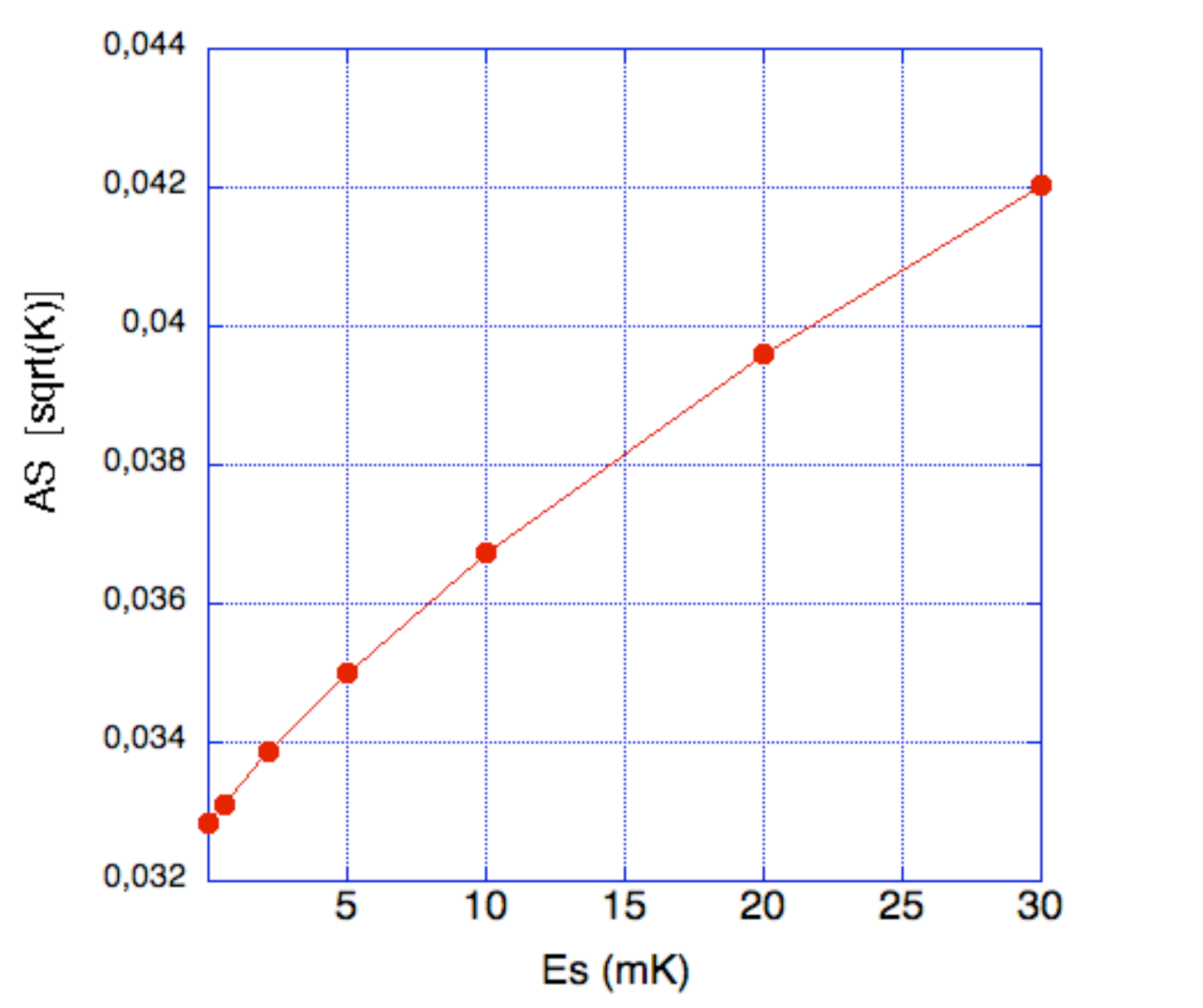}
\caption{
The $AS $ response, in squared $kelvin$ units, versus the energy of cosmic ray calibration signals. Fore each point,  300 off-source measurements are been employed and averaged. At $E_s=0$, the $AS_{bck}$ value is reported. The continuos line is a polynomial fit.
   \label{calibra} }
\end{figure}

\begin{figure}
\includegraphics[width=0.9\linewidth]{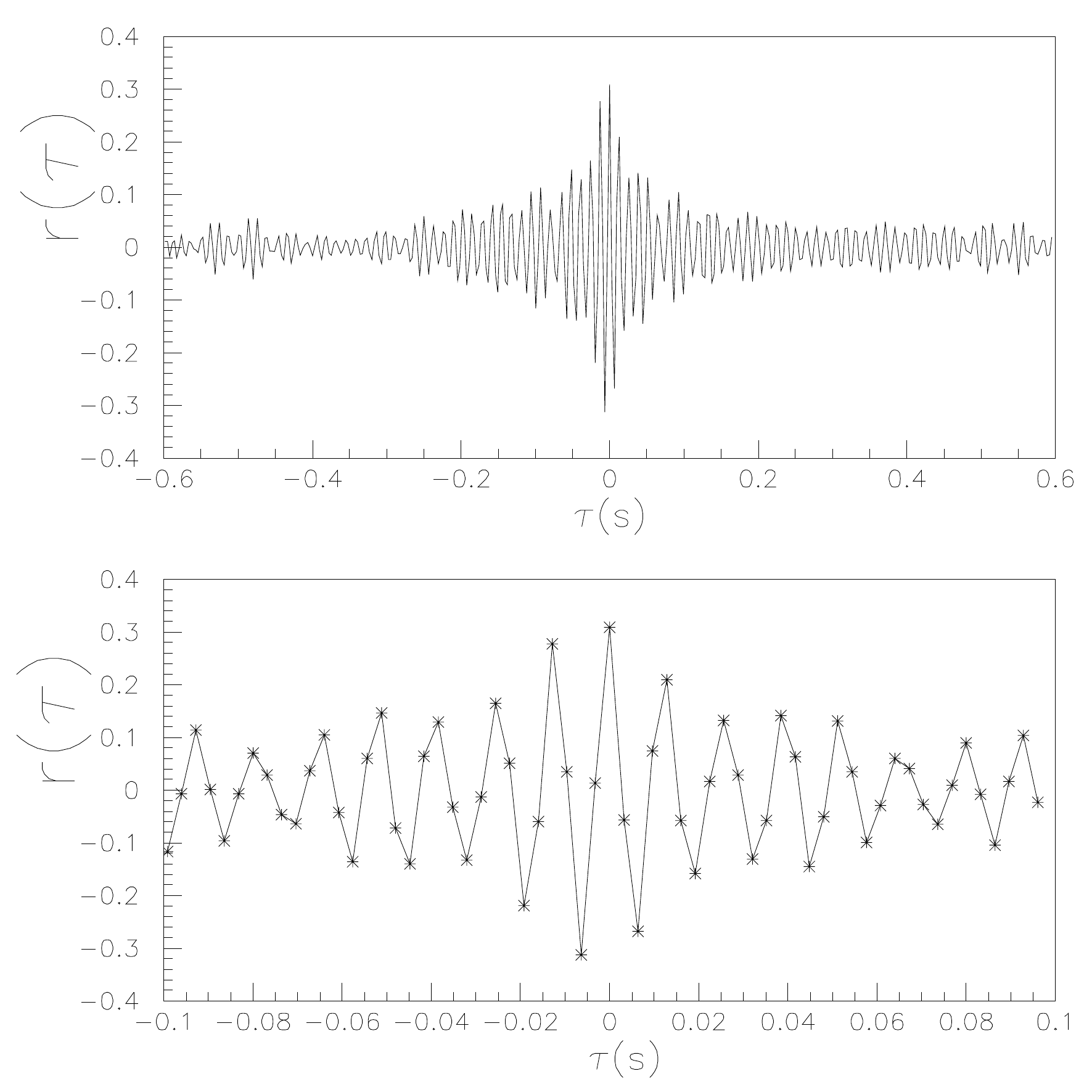}
\caption{
The correlation function of the signals shown in the previous fig.\ref{fig4_cos}, versus the time shift. The $zero$ on the abscissa is arbitrary.
   \label{fig6_cal} }
\end{figure}
\subsection{Cross-correlation response}
\label{cali3}
To study the
cross-correlation response, we use the EASs selected before and the corresponding GW data
stretches showed in fig.\ref{fig4_cos}. The relative cross-correlation
function is shown in fig.\ref{fig6_cal}. We note the oscillating behaviour of the
maximum response envelope in the temporal shift window of a few
ten-milliseconds at zero time shift. This can compromise the efficiency
of the cross-correlaction detection algorithm, especially in very low
signal-to noise-ratio conditions. To counter this, we adopt  the $r_{max}$
parameter, the maximum modulus of the cross-correlation function in the
interval of $\pm$16 ms around the zero shift.

\end{document}